\documentclass[aps,prx,superscriptaddress,twocolumn,nofootinbib,nobibnotes,floatfix,showpacs,reprint,longbibliography]{revtex4-2}
\usepackage{siunitx}
\usepackage{float}
\usepackage[utf8]{inputenc}
\usepackage[T1]{fontenc}
\usepackage{lmodern}
\usepackage{orcidlink}
\usepackage{amsmath,amsfonts,amssymb,amsthm}
\usepackage{bm} 
\usepackage{xcolor} 
\usepackage{xfrac} 
\usepackage{enumitem} 
\usepackage[normalem]{ulem} 
\usepackage{soul} 
\usepackage{subfigure}
\usepackage{multirow}
\usepackage{dcolumn} 
\usepackage{graphicx} 
\usepackage{hyperref} 
\hypersetup{
    unicode={true},
    colorlinks={true},
    linkcolor={blue},
    citecolor={blue},
    urlcolor={blue}
}

\begin{document}
\title{Rare-earth chalcogenide perovskites: A promising class of materials
for optoelectronic applications}
\author{Surajit Adhikari}
\email{sa731@snu.edu.in}

\affiliation{Department of Physics, School of Natural Sciences, Shiv Nadar Institution
of Eminence, Greater Noida, Gautam Buddha Nagar, Uttar Pradesh 201314,
India}
\author{Priya Johari}
\email{priya.johari@snu.edu.in}

\affiliation{Department of Physics, School of Natural Sciences, Shiv Nadar Institution
of Eminence, Greater Noida, Gautam Buddha Nagar, Uttar Pradesh 201314,
India}
\begin{abstract}
Chalcogenide perovskites have attracted significant attention for
optoelectronic applications due to their nontoxic composition, robust
phase stability, and excellent optoelectronic properties. Among them,
rare-earth chalcogenide perovskites have recently emerged as promising
candidates for next-generation devices. However, their excitonic and
polaronic properties remain largely unexplored due to the high computational
cost of accurate theoretical treatments. In this work, we present
a comprehensive first-principles investigation of excitonic dynamics
and polaronic effects in a series of III-III rare-earth chalcogenide
perovskites ABX$_{3}$ (A = Y, La; B = Sc, Y; X = S, Se), along with
their structural stability and optoelectronic properties, using state-of-the-art
density functional theory in conjunction with many-body perturbation
theory within the G$_{0}$W$_{0}$ and Bethe-Salpeter equation (BSE)
frameworks. All investigated compounds satisfy the
dynamical and mechanical stability criteria, although several are
thermodynamically metastable at 0 K and may require kinetic stabilization
and/or finite-temperature effects for experimental realization. They
exhibit quasiparticle band gaps in the range of 2.75$-$4.47 eV, and
the BSE calculations reveal strong optical absorption spanning the
visible to ultraviolet regions. The computed excitonic properties
indicate intermediate-to-large exciton binding energies (0.148$-$0.517
eV), moderately localized excitons, and strong electron-hole
wavefunction overlap, indicative of favorable radiative recombination
characteristics and enhanced light-matter interaction. Furthermore,
analysis based on the Fröhlich model demonstrates intermediate-to-strong
carrier-phonon coupling, with electron-phonon interactions generally
stronger than hole-phonon interactions. Notably, charge-separated
polaronic states are energetically less favorable than bound excitonic
states in most compounds, with the exception of La(Sc, Y)Se$_{3}$,
while hole polarons exhibit significantly higher mobilities (up to
$\sim$ 40 cm$^{2}$V$^{-1}$s$^{-1}$) compared to electron polarons.
Overall, rare-earth chalcogenide perovskites ABX$_{3}$ exhibit a
compelling combination of structural stability, tunable optoelectronic
properties, pronounced excitonic effects, and favorable polaronic
transport, positioning them as promising lead-free materials for next-generation
optoelectronic devices, including light-emitting devices and photodetectors.
\end{abstract}
\maketitle

\section{Introduction:}

Over the last decade, inorganic-organic halide perovskites (IOHPs)
have attracted extraordinary interest owing to their outstanding electronic
and optical characteristics \citep{chapter2-12,chapter3-8,chapter3-9,chapter3-10}.
This intense research focus has led to remarkable progress in perovskite
solar cells, with the power conversion efficiency (PCE) rising rapidly
from an initial value of 3.8\% to a record 27\% \citep{chapter2-12,chapter2-53}.
However, despite these impressive achievements, significant challenges
remain. Most high-performance IOHPs are lead (Pb)-based, giving rise
to serious toxicity concerns. Moreover, their practical applicability
is hindered by poor long-term stability, including degradation and
thermal and chemical instability associated with the organic constituents
\citep{chapter3-12,chapter1-16}. These limitations have motivated
extensive efforts toward the development of environmentally benign,
stable perovskite alternatives, opening new pathways for next-generation
optoelectronic materials.

Recently, chalcogenide perovskites have emerged as a novel class of
functional materials. Owing to their high structural stability, suitable
band gaps, and excellent optoelectronic properties, several members
of this family have been identified as promising candidates for light-absorbing
and light-emitting applications \citep{chapter3-11,chapter3-16,chapter3-17,chapter3-18,chapter3-37}.
Among them, the II-IV chalcogenide perovskites with the general formula
ABX$_{3}$ (where A = Ca, Sr, Ba; B = Ti, Zr, Sn, Hf; and X = S, Se)
are the most extensively studied to date \citep{chapter3-17,chapter3-18,chapter1-63,chapter3-19,chapter5-16,chapter6-9,Ref-PRB}.
Further diversification of this materials family can be achieved by
exploring alternative elemental combinations at the cationic sites.
In this context, the III-III chalcogenide perovskites, although already
synthesized, remain largely unexplored as functional materials \citep{Ref-1,Ref-2,Ref-5,Ref-6}.
For optoelectronic applications, access to a broader materials palette
is highly advantageous, as it enables precise tuning of critical parameters
such as band gaps, band alignments, and structural properties, including
lattice constants required for epitaxial growth. Consequently, the
exploration of III-III chalcogenide perovskites composed of environmentally
benign, non-toxic elements is highly desirable, as these materials
can effectively complement the existing family of II-IV chalcogenide
perovskites.

Recent studies on III-III chalcogenide perovskites have unveiled promising
opportunities for the design of stable and environmentally benign
optoelectronic materials. For instance, Zhang \textit{et al.} theoretically
predicted the orthorhombic $Pnma$ (No. 62) crystal structure of YScS$_{3}$,
reporting an indirect G$_{0}$W$_{0}$ band gap of 3.16 eV \citep{Ref-3}.
Subsequently, through density functional theory calculations, Zhang
\textit{et al.} not only predicted but also experimentally synthesized
a light-emitting III-III-S$_{3}$ perovskite, LaScS$_{3}$, exhibiting
a direct band gap of 2.62 eV \citep{Ref-1}. Furthermore, they also
reported the prediction and experimental realization of the direct
band gap material LaScSe$_{3}$, with a G$_{0}$W$_{0}$ band gap
of 2.96 eV \citep{Ref-2}. In addition, the $P2_{1}/m$ crystal structure
of the LaYS$_{3}$ perovskite has been both theoretically predicted
and experimentally synthesized, exhibiting an optimal band gap of
2.0 eV, making it a promising wide band gap photoabsorber for tandem
solar energy conversion devices \citep{Ref-5,Ref-6}.

Most of the aforementioned studies have primarily focused on the electronic
and optical properties of III-III rare-earth chalcogenide perovskites.
In contrast, their excitonic and polaronic properties remain largely
unexplored, despite their critical role in determining the performance
of optoelectronic devices. Excitonic effects govern fundamental processes
such as light absorption, emission, and charge separation. In contrast,
polaron formation, arising from carrier-phonon interactions, strongly
influences charge transport and carrier mobility \citep{chapter5-16,chapter6-9,Ref-PRB}.
A quantitative description of these properties, however, is computationally
demanding and has therefore received limited attention. This gap highlights
the need for a comprehensive investigation of excitonic and polaronic
effects in these compounds, which has not yet been achieved and is
addressed in the present work.

In this work, we undertake a systematic and comprehensive investigation
of the structural stability, as well as the electronic, optical, excitonic,
and polaronic properties of III-III rare-earth chalcogenide perovskites
ABX$_{3}$ (A = Y, La; B = Sc, Y; X = S, Se) in the orthorhombic $Pnma$
phase, a widely observed and energetically competitive distorted perovskite
structure that provides a consistent framework for comparative analysis.
Our study is based on first-principles calculations within the frameworks
of density functional theory (DFT) \citep{chapter2-36,chapter2-37},
density functional perturbation theory (DFPT) \citep{chapter1-60},
and many-body perturbation theory (MBPT) \citep{chapter3-1,chapter3-2}.
Initially, the crystal structures are optimized using the semilocal
PBE \citep{chapter1-34} as well as PBEsol \citep{Ref-7} exchange-correlation
(xc) functionals, and all compounds are found to be both dynamically
and mechanically stable. However, several are thermodynamically
metastable at 0 K, suggesting that their experimental realization
may require kinetic stabilization and/or finite-temperature effects. Subsequently, the electronic properties are investigated using both
the HSE06 hybrid xc functional \citep{chapter1-35} and the G$_{0}$W$_{0}$@PBE
approach \citep{chapter1-69,chapter1-70}, revealing quasiparticle
band gaps in the range of 2.75$-$4.47 eV. Following that, the optical
response is computed by solving the Bethe-Salpeter equation (BSE)
\citep{chapter1-67,chapter1-68} on top of G$_{0}$W$_{0}$@PBE, enabling
the evaluation of the dielectric function and exciton binding energies.
The results indicate that these perovskites exhibit intermediate-to-high
exciton binding energies, moderately localized excitons, and strong
electron-hole wavefunction overlap, suggesting favorable radiative
transitions and pronounced light-matter interaction. Finally, we
investigate the effects of carrier-phonon coupling and estimate the
polaron mobility using the Fröhlich model \citep{chapter2-51} and
the Hellwarth polaron model \citep{chapter2-22}. The analysis reveals
intermediate to strong carrier-phonon coupling, with electron-phonon
interactions being more pronounced than those of holes, resulting
in lower electron mobility and comparatively higher hole mobility.
Overall, this work provides a comprehensive understanding of rare-earth
ABX$_{3}$ (A = Y, La; B = Sc, Y; X = S, Se) chalcogenide perovskite
compounds, highlighting their potential as lead-free perovskites with
promising optoelectronic properties for next-generation devices.

\section{Computational Details:}

In this work, first-principles calculations were carried out within
the framework of density functional theory (DFT) \citep{chapter2-36,chapter2-37},
density functional perturbation theory (DFPT) \citep{chapter1-60},
and many-body perturbation theory (MBPT) \citep{chapter3-1,chapter3-2},
as implemented in the Vienna ab initio Simulation Package (VASP) \citep{chapter1-31,chapter1-32}.
The interaction between valence electrons and ionic cores was described
using projector augmented-wave (PAW) pseudopotentials \citep{chapter1-33}.
Structural optimizations were carried out within the generalized gradient
approximation (GGA) using the Perdew-Burke-Ernzerhof (PBE) \citep{chapter1-34}
and the PBEsol \citep{Ref-7} exchange-correlation (xc) functionals.
The inclusion of PBEsol is motivated by its improved accuracy in describing
equilibrium properties of solids, particularly lattice constants and
structural distortions, which are crucial for reliably assessing dynamical
stability and phonon properties. A plane-wave cutoff energy of 400
eV was employed, and the electronic self-consistent field convergence
criterion was set to $10^{-6}$ eV. All structures were fully optimized
until the Hellmann-Feynman forces on each atom were less than 0.01
eV/$\textrm{\AA}$. Brillouin zone integrations were carried out using
a $\Gamma$-centered $7\times7\times5$ k-point mesh. The optimized
crystal structures were visualized using the VESTA package \citep{chapter2-3}.
Phonon dispersion curves were calculated using DFPT as implemented
in the PHONOPY package \citep{chapter3-6}, employing $2\times2\times2$
supercells. To assess the functional dependence of
the dynamical stability, the phonon spectra were computed for crystal
structures relaxed using both the PBE and PBEsol exchange-correlation
functionals.

Electronic band structures
were initially obtained using the PBE xc functional including spin-orbit
coupling (SOC), which was found to have a negligible impact on the
overall band dispersion. To achieve improved accuracy in bandgap estimation,
hybrid xc functional calculations using HSE06 \citep{chapter1-35}
and quasiparticle corrections within the G$_{0}$W$_{0}$@PBE \citep{chapter1-69,chapter1-70}
approach were performed. A $\Gamma$-centered $3\times3\times2$
k-point mesh was employed for the G$_{0}$W$_{0}$ calculations. This
choice is justified by the demonstrated convergence of the quasiparticle
band gaps, while the use of denser k-point meshes for the present
20-atom unit cells would incur a substantially higher computational
cost and is therefore beyond the scope of the present work (for details,
see Sec. X of the SM \citep{supp}). The convergence of the quasiparticle
bandgap with respect to the number of unoccupied bands (NBANDS), the
plane-wave cutoff energy (ENCUT), and the response-function cutoff
energy (ENCUTGW) was carefully examined (for details, see Sec. X of
the SM \citep{supp}). Based on these convergence tests, 640 bands,
a plane-wave cutoff energy of 400 eV, and a response-function cutoff
energy of 300 eV were adopted, yielding quasiparticle bandgaps converged
to within 0.05 eV. Carrier effective masses were calculated using
the SUMO code \citep{chapter2-10} via parabolic fitting near the
band extrema. Optical properties were further refined by solving the
Bethe-Salpeter equation (BSE) \citep{chapter1-67,chapter1-68} on
top of G$_{0}$W$_{0}$@PBE, explicitly including electron-hole interactions. The BSE kernel was constructed using 24 occupied and
24 unoccupied bands, based on convergence tests (for details, see
Sec. XI of the SM \citep{supp}). Elastic and optical properties
were post-processed using the VASPKIT package \citep{chapter1-48},
and the ionic contribution to the dielectric constant was obtained
from DFPT calculations.

\section{Results and Discussions:}

In the present study, we conduct a comprehensive investigation of
the optoelectronic properties of III-III rare-earth chalcogenide perovskites
ABX$_{3}$ (A = Y, La; B = Sc, Y; X = S, Se). The following sections
provide an in-depth analysis of their structural stability, electronic
structure, and transport characteristics, along with their optical
response, excitonic behavior, and polaronic effects. These results
aim to establish a fundamental understanding of these materials and
offer valuable insights to guide future experimental studies.

\begin{figure}[H]

\begin{centering}
\includegraphics[width=0.48\textwidth,height=0.75\textheight,keepaspectratio]{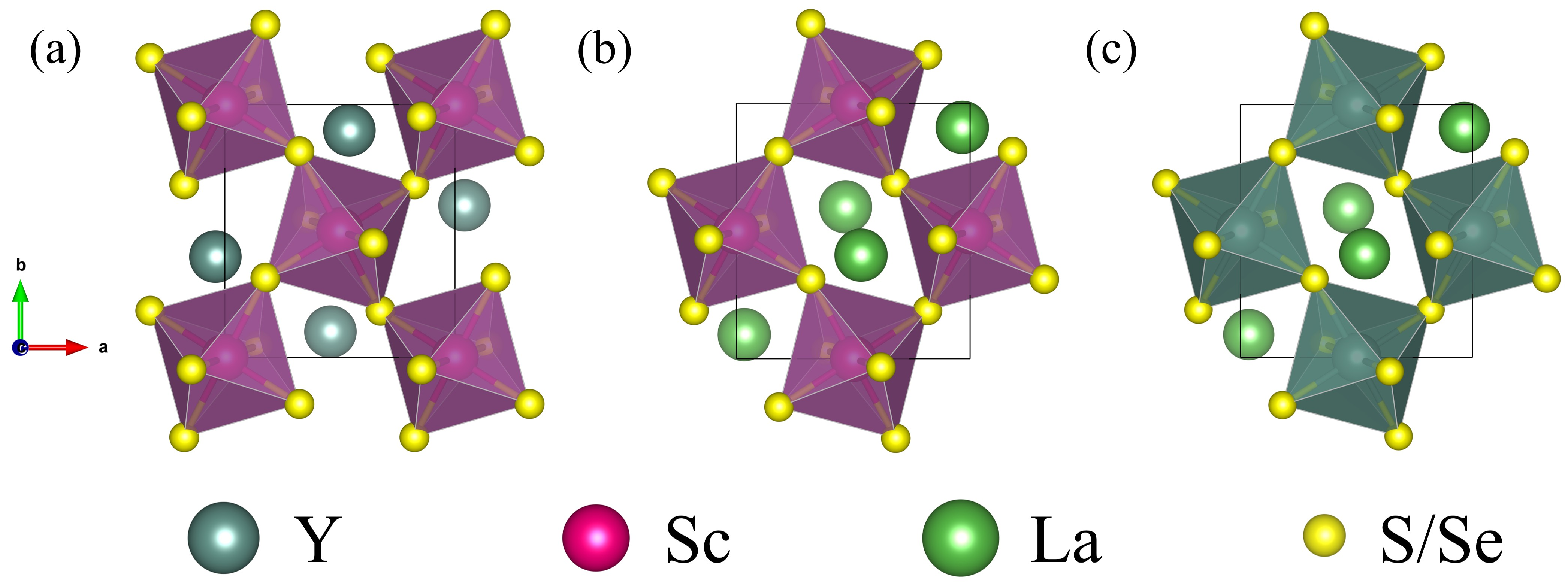}\caption{\label{fig:1}Crystal structures of rare-earth chalcogenide perovskites
(a) YScX$_{3}$, (b) LaScX$_{3}$, and (c) LaYX$_{3}$, respectively,
where X = S, Se.}
\par\end{centering}
\end{figure}

\subsection{Structural Properties:}

Figure \ref{fig:1} illustrates the orthorhombic crystal structures
of rare-earth chalcogenide perovskites ABX$_{3}$ (A = Y, La; B =
Sc, Y; and X = S, Se), crystallizing in the $Pnma$ space group (No.
62). Owing to the +3 valency of both A- and B-site cations, these
materials are categorized as III-III chalcogenide perovskites. The
unit cell contains four formula units (20 atoms), comprising 4 A-site
cations (Y or La), 4 B-site cations (Sc or Y), and 12 chalcogen anion
atoms (S or Se). In this structure, the A-site cations exhibit 12-fold
coordination, forming cuboctahedral environments with the surrounding
chalcogen atoms. In contrast, the B-site cations are 6-fold coordinated,
giving rise to corner-sharing distorted $\mathrm{[BX_{6}]^{9-}}$
octahedra. These octahedra are both tilted and distorted, leading
to the characteristic orthorhombic symmetry of the $Pnma$ phase \citep{chapter3-37}.
In this study, we report for the first time the $Pnma$ phase of YScSe$_{3}$,
LaYS$_{3}$, and LaYSe$_{3}$ compounds. The lattice parameters of
the optimized structures are calculated using both PBE and PBEsol
xc functionals and are summarized in Table \ref{tab:1}. It is found
that the lattice parameters of LaScSe$_{3}$ are in good agreement
with the available experimental results (a = 6.77 $\textrm{\AA}$, b
= 7.53 $\textrm{\AA}$, and c = 10.00 $\textrm{\AA}$ \citep{Ref-2}).
Moreover, the PBE functional slightly overestimates the lattice parameters,
while PBEsol slightly underestimates them, with both deviations being
of comparable magnitude. This indicates that both functionals yield
similar accuracy, despite exhibiting opposite systematic deviations.
In addition, the octahedral distortion parameters, including the average
bond length, polyhedral volume, bond angle variance, and bond-length
distortion index, for the BX$_{6}$ octahedra in these chalcogenide
perovskites are computed using both functionals and presented in Table
S1 (for details, see Sec. I of the SM \citep{supp}).

\begin{table*}[t]
\caption{\label{tab:1}Calculated lattice parameters and decomposition energies
($\mathrm{\Delta H_{D}}$) of ABX$_{3}$ (A = Y, La; B = Sc, Y; and
X = S, Se) rare-earth chalcogenide perovskites using the PBE and PBEsol
xc functional, respectively (PBEsol values in bold).}

\centering{}%
\begin{tabular}{ccccccccc}
\hline 
\multirow{2}{*}{Configurations} &  &  & \multicolumn{3}{c}{Lattice parameters ($\textrm{\AA}$)} &  &  & $\mathrm{\Delta H_{D}}$\tabularnewline
\cline{4-6} \cline{5-6} \cline{6-6} 
 &  &  & a & b & c &  &  & (meV/atom)\tabularnewline
\hline 
YScS$_{3}$ &  &  & 6.39 (\textbf{6.27}) & 7.02 (\textbf{6.97}) & 9.55 (\textbf{9.37}) &  &  & -9.1 (\textbf{5.0})\tabularnewline
YScSe$_{3}$ &  &  & 6.67 (\textbf{6.54}) & 7.35 (\textbf{7.29}) & 10.00 (\textbf{9.79}) &  &  & -28.9 (\textbf{-6.8})\tabularnewline
LaScS$_{3}$ &  &  & 6.59 (\textbf{6.46}) & 7.21 (\textbf{7.16}) & 9.65 (\textbf{9.51}) &  &  & 26.0 (\textbf{26.1})\tabularnewline
\multirow{1}{*}{LaScSe$_{3}$} &  &  & 6.83 (\textbf{6.68}) & 7.58 (\textbf{7.51}) & 10.08 (\textbf{9.92}) &  &  & 13.5 (\textbf{8.8})\tabularnewline
LaYS$_{3}$ &  &  & 6.81 (\textbf{6.70}) & 7.42 (\textbf{7.37}) & 10.06 (\textbf{9.88}) &  &  & -2.7 (\textbf{-19.9})\tabularnewline
LaYSe$_{3}$ &  &  & 7.04 (\textbf{6.90}) & 7.78 (\textbf{7.72}) & 10.48 (\textbf{10.26}) &  &  & -15.2 (\textbf{-25})\tabularnewline
\hline 
\end{tabular}
\end{table*}

To assess the thermodynamic stability against decomposition, we calculated
the decomposition enthalpy according to

\begin{equation}
\mathrm{ABX_{3}\rightarrow1/2\:A_{2}X_{3}+1/2\:B_{2}X_{3}}
\end{equation}

for these III-III-X$_{3}$ compounds that pass the phase stability
criteria (for details, see Sec. II of the SM \citep{supp}). The decomposition
enthalpy ($\mathrm{\Delta H_{D}}$) of these compounds is evaluated
using both PBE and PBEsol xc functionals and is tabulated in Table
\ref{tab:1}. The LaScX$_{3}$ (X = S, Se) compounds exhibit positive
$\mathrm{\Delta H_{D}}$ values with both functionals, indicating
thermodynamic stability with respect to the considered
decomposition pathway. In contrast, the remaining
compounds exhibit negative $\mathrm{\Delta H_{D}}$ values, indicating
thermodynamic metastability at 0 K with respect to this decomposition
pathway. The only exception is YScS$_{3}$, which yields a positive
$\mathrm{\Delta H_{D}}$ when using the PBEsol functional. Nevertheless, decomposition enthalpy based on a single
reaction does not constitute a complete thermodynamic stability criterion,
since alternative competing phases and decomposition pathways may
exist. A rigorous assessment therefore requires construction of the
convex-hull phase diagram, which compares the energy of a compound
against all known competing phases within the corresponding chemical
space. Compounds lying on the convex hull are thermodynamically stable,
whereas those with positive energy above the hull ($\mathrm{E_{hull}}$)
are metastable with respect to decomposition into a combination of
lower-energy phases.

To directly address this limitation, we constructed
the convex-hull phase diagrams for each A$-$B$-$X (A = Y, La; B
= Sc, Y; and X = S, Se) chemical space by incorporating all relevant
competing binary and ternary phases at the same DFT level. The calculated
$\mathrm{E_{hull}}$ values are summarized in Table S3, while the
corresponding convex-hull diagrams are presented in Figure S1 (for
details, see Sec. III of the SM \citep{supp}). The results show that
LaScX$_{3}$ (X = S, Se) lie on the convex hull with $\mathrm{E_{hull}}$
= 0, confirming their thermodynamic stability at 0 K. The remaining
compounds possess finite positive $\mathrm{E_{hull}}$ values (0.009$-$0.029
eV), indicating metastability with respect to competing equilibrium
phases. It is important to note, however, that a positive energy above
the hull does not necessarily preclude experimental realization. Numerous
experimentally synthesized materials are known to exist as metastable
phases because finite-temperature vibrational and configurational
entropy, kinetic barriers, and non-equilibrium synthesis routes can
effectively stabilize compounds that are slightly above the convex
hull \citep{ayan}. Consequently, the above-hull compounds identified
here should be regarded as metastable candidates that may still be
experimentally accessible under suitable synthesis conditions rather
than as equilibrium-stable phases. However, thermodynamic stability
alone is insufficient to ensure stability of these materials; therefore,
dynamical and mechanical stability are also examined.

\begin{figure}[H]
\begin{centering}
\includegraphics[width=0.48\textwidth,height=1\textheight,keepaspectratio]{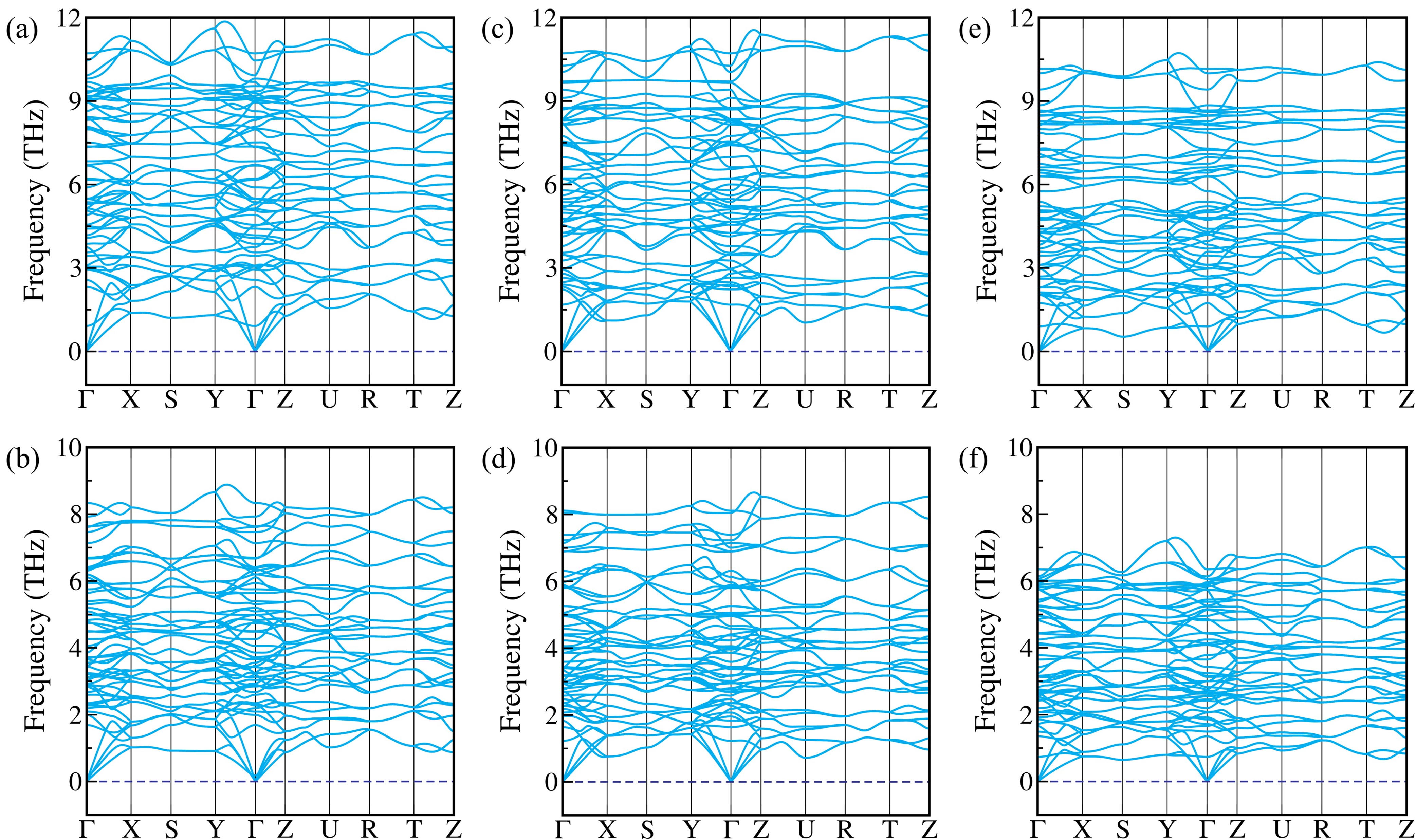}
\par\end{centering}
\caption{\label{fig:2}Phonon dispersion curves of the rare-earth chalcogenide
perovskites (a) YScS$_{3}$, (b) YScSe$_{3}$, (c) LaScS$_{3}$, (d)
LaScSe$_{3}$, (e) LaYS$_{3}$, and (f) LaYSe$_{3}$, computed using
the DFPT method based on PBEsol-relaxed structures.}

\end{figure}

The dynamical stability of the investigated perovskites serves as
a crucial indicator of their structural integrity and suitability
for functional applications. It is evaluated through phonon dispersion
relations, which describe the vibrational behavior of atoms in the
crystal lattice. For a dynamically stable system at 0 K, all phonon
frequencies are required to be real and positive across the entire
Brillouin zone; the occurrence of imaginary (negative) frequencies
signifies possible structural instabilities. To assess this, self-consistent
phonon calculations are performed within the DFPT framework using
the structures relaxed with both PBE and PBEsol xc functionals. The
computed phonon dispersion curves of ABX$_{3}$ (A = Y, La; B = Sc,
Y; and X = S, Se), obtained using structures relaxed with the PBE
and PBEsol xc functionals, are presented in Figs. S2 and \ref{fig:2},
respectively. Within the PBE functional, YScS$_{3}$, LaScS$_{3}$,
LaYS$_{3}$, and LaYSe$_{3}$ are dynamically stable at 0 K, as confirmed
by the absence of imaginary phonon modes, whereas (Y, La)ScSe$_{3}$
shows dynamical instabilities, as indicated by the presence of imaginary
modes. Specifically, YScSe$_{3}$ shows a maximum
imaginary frequency of -0.333 THz at the $\Gamma$ point, while
LaScSe$_{3}$ exhibits imaginary frequencies of approximately -0.189
THz along the $\Gamma$$-$X direction and -0.349 THz at the U point.
The relatively small magnitudes of these imaginary frequencies suggest
the presence of soft lattice instabilities rather than pronounced
structural instabilities. This interpretation is further supported
by the experimental realization of LaScSe$_{3}$, indicating that
these weak instabilities are unlikely to preclude its synthesis and
may instead reflect the sensitivity of the lattice dynamics to the
equilibrium structural parameters \citep{Ref-2}. In contrast, phonon
calculations based on the PBEsol-relaxed structures show no imaginary
phonon modes throughout the Brillouin zone for any of the investigated
compounds, indicating dynamical stability within the PBEsol framework.
Together with the PBE results, these findings suggest that the lattice
dynamics of some compounds are sensitive to small variations in the
equilibrium lattice parameters and are therefore dependent on the choice of exchange-correlation functional.

Beyond the thermodynamic and dynamical stability discussed above,
the mechanical stability and corresponding elastic properties of ABX$_{3}$
(A = Y, La; B = Sc, Y; X = S, Se) chalcogenide perovskites are further
examined. It is well known that the suitability of a material for
practical device applications is strongly influenced by its elastic
behavior. The second-order elastic constants ($C_{ij}$) are calculated
using the energy-strain method \citep{chapter1-47}, from which the
relevant elastic properties are derived (for details, see Sec. V of
the SM \citep{supp}). Owing to the orthorhombic symmetry of all considered
compounds, nine independent elastic constants ($C_{11}$, $C_{22}$,
$C_{33}$, $C_{44}$, $C_{55}$, $C_{66}$, $C_{12}$, $C_{13}$,
and $C_{23}$) are sufficient to describe their mechanical stability
and elastic behavior. The calculated $C_{ij}$ values, listed in Table
S4, satisfy the Born stability criteria \citep{chapter1-47}, thereby
confirming the excellent mechanical stability of these orthorhombic
chalcogenide perovskites.

Further, the bulk modulus ($B$), shear modulus ($G$), Young\textquoteright s
modulus ($Y$), and Poisson\textquoteright s ratio ($\nu$) are evaluated
within the Voigt-Reuss-Hill approximation \citep{chapter1-49,chapter1-50}
and are summarized in Table S4. The relatively larger values of $B$
compared to $G$ indicate that the investigated chalcogenide perovskites
exhibit greater resistance to volume change than to shear deformation.
The comparatively lower values of $G$ and $Y$ further suggest that
these materials possess a certain degree of mechanical flexibility.
To assess the ductile or brittle nature of these systems, we employ
Pugh\textquoteright s criterion based on the $B/G$ ratio \citep{chapter1-51},
along with Poisson\textquoteright s ratio ($\nu$). The obtained values
of $B/G>1.75$ and $\nu>0.26$ consistently indicate that all the
compounds exhibit ductile behavior. These characteristics suggest
that the investigated materials are mechanically robust yet sufficiently
flexible for potential device applications.

Overall, the synthesizability of these III-III chalcogenide perovskites
is established by the combined assessment of thermodynamic, dynamical,
and mechanical stability. Notably, LaScX$_{3}$ (X = S, Se), which
has already been experimentally synthesized \citep{Ref-1,Ref-2},
is found to be thermodynamically stable, consistent with its positive
decomposition enthalpy. In contrast, the remaining compounds are metastable
with respect to decomposition. Nevertheless, the absence of imaginary
phonon modes (within PBEsol) and the satisfaction of mechanical stability
criteria confirm their dynamical and mechanical stability. These results
suggest that, despite thermodynamic metastability, the compounds are
likely experimentally accessible through kinetic stabilization or
finite-temperature effects, with LaScX$_{3}$ serving as a benchmark
system validating the reliability of the present approach.

\subsection{Electronic properties:}

Following the assessment of structural stability, the electronic properties,
such as the band structure and partial density of states (PDOS) of
rare-earth chalcogenide perovskites ABX$_{3}$ (A = Y, La; B = Sc,
Y; and X = S, Se) are evaluated to gain fundamental insight into their
suitability for optoelectronic applications. At first, the electronic
band structures of these chalcogenide perovskites are calculated using
the semilocal GGA-PBE xc functional, both with and without inclusion
of spin-orbit coupling (SOC). The GGA-PBE xc functional is found to
underestimate the band gaps due to the self-interaction error of electrons
(see Table \ref{tab:2}). Furthermore, SOC has a negligible effect
on the band gap (see Table S5 of the SM \citep{supp}), as expected
for chalcogenide perovskites \citep{chapter5-16,chapter6-9,Ref-PRB}.

\begin{figure}[H]
\begin{centering}
\includegraphics[width=0.48\textwidth,height=1\textheight,keepaspectratio]{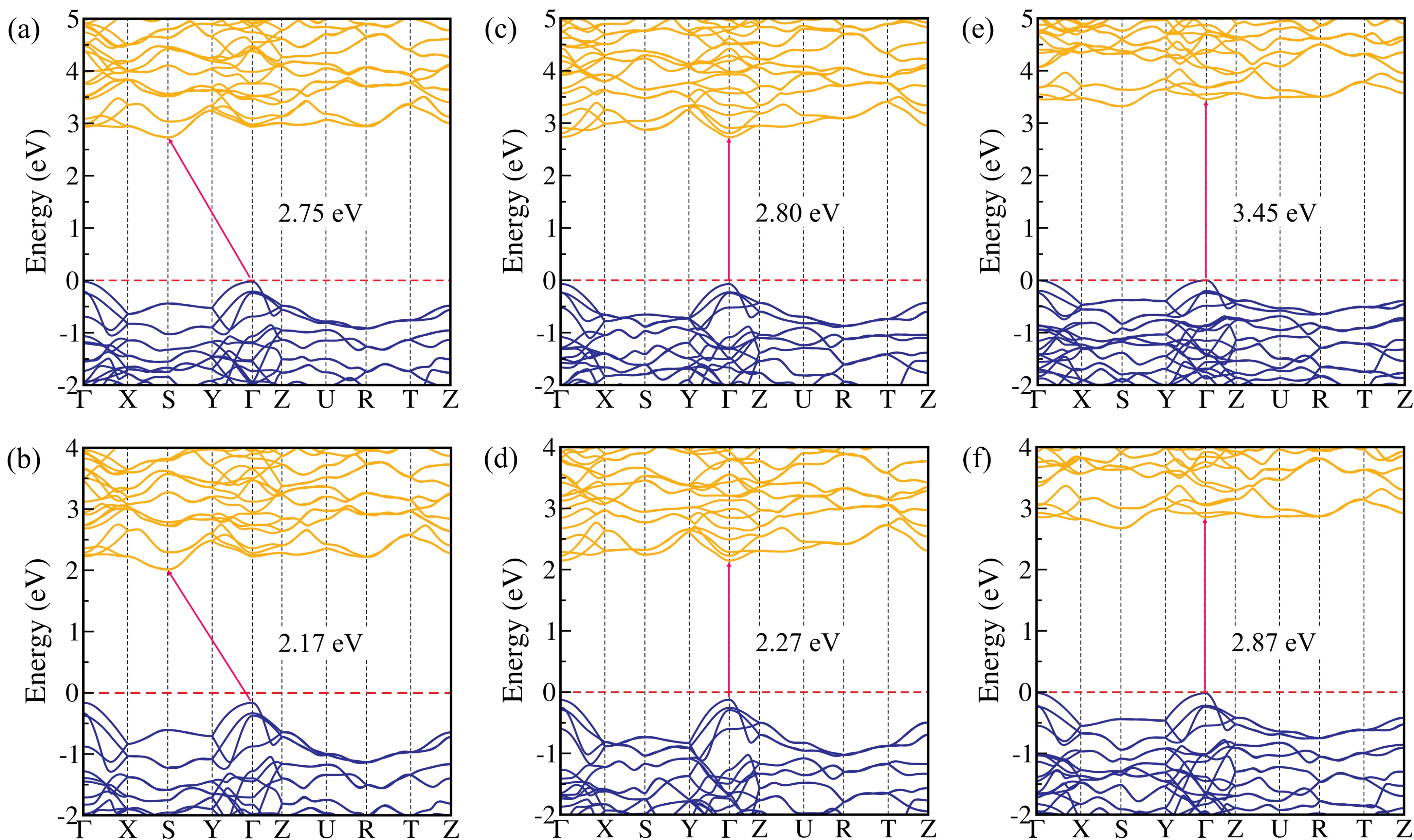}
\par\end{centering}
\caption{\label{fig:3}Electronic band structures of the rare-earth chalcogenide
perovskites (a) YScS$_{3}$, (b) YScSe$_{3}$, (c) LaScS$_{3}$, (d)
LaScSe$_{3}$, (e) LaYS$_{3}$, and (f) LaYSe$_{3}$, calculated using
the HSE06 xc functional. The Fermi level is set to be zero and marked
by the dashed line.}
\end{figure}

Subsequently, the band gaps are computed more accurately using the
hybrid HSE06 xc functional and the many-body perturbation theory (MBPT)-based
GW approach, specifically at the G$_{0}$W$_{0}$@PBE level. Figure
\ref{fig:3} presents the band structures of these compounds, calculated
using the hybrid HSE06 functional. Our calculated electronic band
structures explicitly demonstrate that YScX$_{3}$ (X = S, Se) are
indirect band gap materials. In these compounds, the valence band
maximum (VBM) is located at the $\Gamma$-point, whereas the conduction
band minimum (CBM) occurs at the $\mathrm{S}$-point in the Brillouin
zone {[}see Fig. \ref{fig:3}(a)-(b){]}. This spatial separation of
the band extrema confirms the indirect nature of the fundamental band
gap, implying that optical transitions near the band edge are phonon-assisted.
In contrast, all other compounds investigated in this study exhibit
direct band gaps. For these systems, both the VBM and CBM are located
at the same high-symmetry $\mathrm{k}$-point, specifically at the
$\Gamma$-point {[}see Fig. \ref{fig:3}(c)-(f){]}. As a result, the
fundamental electronic transition is momentum-conserving, leading
to a stronger optical absorption onset compared to the indirect-gap
counterparts.

The band gaps of these compounds, as calculated using the HSE06 functional
and the G$_{0}$W$_{0}$@PBE approach, are summarized in Table \ref{tab:2}.
The HSE06 band gaps of these chalcogenide perovskites are found to
lie in the range of 2.17$-$3.45 eV, whereas the G$_{0}$W$_{0}$@PBE
band gaps are systematically larger, spanning 2.75$-$4.47 eV. This
trend reflects the well-known tendency of many-body perturbation theory
within the GW approximation to yield improved quasiparticle energies
by explicitly accounting for electron-electron interactions, thereby
correcting the band gap underestimation inherent to standard and hybrid
density functional approaches. Importantly, the G$_{0}$W$_{0}$@PBE
results obtained in this work are in good agreement with previously
reported theoretical studies \citep{Ref-2,Ref-3}, confirming the
reliability of the present computational framework. The absence of
exact quantitative agreement with previous reports can be attributed
to the sensitivity of G$_{0}$W$_{0}$ calculations to the choice
of starting functional, convergence parameters, and structural inputs. For the benchmark compound LaScS$_{3}$, the experimentally
measured band gap (2.62 eV) is closer to the HSE06 value (2.80 eV)
than to the G$_{0}$W$_{0}$@PBE value (3.72 eV) \citep{Ref-1}. This
discrepancy is likely attributable to the inherent starting-point
dependence of the single-shot G$_{0}$W$_{0}$ approximation, the
absence of self-consistency in the GW calculations, residual convergence
limitations, and the neglect of finite-temperature and excitonic effects. Although G$_{0}$W$_{0}$ calculations based on an
HSE06 starting point or self-consistent GW schemes could further reduce
the starting-point dependence and improve the quantitative accuracy
of the quasiparticle energies, such calculations are computationally
prohibitive for the large unit cells considered here. Nevertheless,
the good agreement with previous theoretical studies supports the
reliability of the present G$_{0}$W$_{0}$ results. Overall, the
calculated band gaps place these chalcogenide perovskites in the visible-to-ultraviolet
wide-band-gap regime, highlighting their potential for optoelectronic
applications. Direct-gap compounds favor efficient light absorption
and emission (e.g., LEDs), whereas indirect-gap systems are advantageous
for longer carrier lifetime\textendash based applications such as
photodetectors and photocatalysis.

\begin{table*}[t]
\caption{\label{tab:2}Bandgap (in eV) of ABX$_{3}$ (A = Y, La; B = Sc, Y;
and X = S, Se) rare-earth chalcogenide perovskites calculated using
different methods as well as computed average effective mass of electron
($m_{e}^{*}$) and hole ($m_{h}^{*}$). Here, $i$, $d$, $t$, and
$e$ represent indirect, direct, theoretical, and experimental bandgaps,
respectively. All values of the effective mass are in terms of free-electron
mass ($m_{0}$).}

\centering{}%
\begin{tabular}{ccccccc}
\hline 
\multirow{1}{*}{Configurations} & PBE & HSE06 & G$_{0}$W$_{0}$@PBE & Previous Work & $m_{e}^{*}$ & $m_{h}^{*}$\tabularnewline
\hline 
YScS$_{3}$ & 1.76$^{i}$ (1.88$^{d}$) & 2.75$^{i}$ (2.96$^{d}$) & 3.49$^{i}$ (3.75$^{d}$) & 3.16$_{t}$ (G$_{0}$W$_{0}$) \citep{Ref-3} & 1.545 & 0.494\tabularnewline
YScSe$_{3}$ & 1.30$^{i}$ (1.43$^{d}$) & 2.17$^{i}$ (2.39$^{d}$) & 2.75$^{i}$ (3.00$^{d}$) &  & 1.622 & 0.375\tabularnewline
LaScS$_{3}$ & 1.69 & 2.80 & 3.72 & 2.62$_{e}$ \citep{Ref-1} & 0.916 & 0.457\tabularnewline
LaScSe$_{3}$ & 1.27 & 2.27 & 2.94 & 2.96$_{t}$ (G$_{0}$W$_{0}$) \citep{Ref-2} & 0.851 & 0.350\tabularnewline
LaYS$_{3}$ & 2.25 & 3.45 & 4.47 &  & 1.255 & 0.542\tabularnewline
LaYSe$_{3}$ & 1.79 & 2.87 & 3.70 &  & 1.180 & 0.419\tabularnewline
\hline 
\end{tabular}
\end{table*}

To gain deeper insight into the electronic structure, the density
of states (DOS) is calculated using the HSE06 xc functional, as shown
in Fig. S3 of the SM \citep{supp}. For YScS$_{3}$, the VBM is predominantly
derived from S-3$p$ orbitals, whereas the CBM is mainly contributed
by Sc-3$d$ orbitals. As expected, the CBM also exhibits hybridization
between Y-4$d$ and Sc-3$d$ states. Similarly, in YScSe$_{3}$, the
VBM is primarily composed of Se-4$p$ orbitals, while the CBM shows
a noticeable downward shift due to enhanced hybridization between
Se-4$p$ and transition-metal $d$ states, leading to the observed
reduction in the band gap. In contrast, for LaScX$_{3}$ (X = S, Se),
the CBM arises from hybridized La-5$d$ and Sc-3$d$ states. The direct-to-indirect
band gap transition primarily arises from the interplay between $d$-$d$
orbital hybridization and structural distortions. In YScX$_{3}$,
stronger Y-4$d$$-$Sc-3$d$ coupling and larger octahedral tilting
stabilize the CBM at the $\mathrm{S}$-point, resulting in an indirect
gap. In contrast, weaker La-5$d$$-$Sc-3$d$ coupling and reduced
tilting in LaScX$_{3}$ shift the conduction band upward at $\mathrm{S}$-point
and favor a minimum at $\Gamma$-point, leading to a direct band gap
\citep{Ref-1}. Furthermore, upon replacing Sc with Y (LaScX$_{3}$
$\rightarrow$ LaYX$_{3}$), the direct band gap nature is preserved,
while the band gap increases due to the higher energy and modified
dispersion of Y-4$d$ states relative to Sc-3$d$ states, resulting
in an upward shift of the CBM in LaYX$_{3}$.

To further substantiate the above interpretation,
we have included orbital-projected band structures near the VBM and
CBM in Fig. S4 of the SM \citep{supp}. These results explicitly identify
the orbital character of the band-edge states and confirm that the
evolution from a direct to an indirect band gap is governed by changes
in the hybridization between Y-4$d$/La-5$d$ and Sc-3$d$ orbitals,
together with the accompanying structural distortions. The orbital-projected
bands provide direct microscopic evidence for the orbital interactions
responsible for the shift of the CBM between the $\Gamma$ and $\mathrm{S}$
points, thereby supporting the mechanism proposed from the DOS analysis.

To gain insight into charge-carrier transport, we further calculate
the effective masses of electrons ($m_{e}^{*}$) and holes ($m_{h}^{*}$)
for all compounds by fitting the $E-k$ dispersion
obtained from PBE band structures using the formula, $\text{\ensuremath{m^{*}=\hbar^{2}\left[\partial^{2}E(k)/\partial k^{2}\right]^{-1}}}$,
and the results are summarized in Table \ref{tab:2}. For the direct
band gap compounds, the effective masses are evaluated as the harmonic
mean along the $\Gamma-$X, $\Gamma-$Y,
and $\Gamma-$Z directions, with results
listed in Table S6 of the SM \citep{supp}. In contrast, for YScX$_{3}$
(X = S, Se), the electron effective masses are obtained by averaging
along the S$-$X and S$-$Y directions, consistent with the location of the CBM. From Table \ref{tab:2},
it is evident that $m_{e}^{*}$ is significantly larger than $m_{h}^{*}$
in these compounds, indicating that holes are expected to possess
higher mobility than electrons. This suggests that these materials
are more favorable for p-type transport.

\subsection{Optical Properties:}

Beyond the electronic structure, the optical response provides key
insight into the suitability of these materials for optoelectronic
applications. In this work, the optical response is accurately described
by capturing many-body effects through the Bethe-Salpeter equation
(BSE) built upon G$_{0}$W$_{0}$@PBE quasiparticle energies, ensuring
a reliable treatment of excitonic contributions to the optical spectra.
In this framework, GW calculations determine the fundamental bandgap,
which is directly comparable to photoelectron (PES) and inverse photoelectron
spectroscopy (IPES) \citep{chapter1-69,chapter1-70}, whereas BSE
yields the optical bandgap consistent with experimental absorption
measurements \citep{chapter1-67,chapter1-68}.

To evaluate the optical response of ABX$_{3}$ (A = Y, La; B = Sc,
Y; and X = S, Se), we focus on the frequency-dependent complex dielectric
function, $\varepsilon(\omega)$ = {[}Re($\varepsilon$){]} + i{[}Im($\varepsilon$){]},
which governs the interaction of electromagnetic radiation with the
material. The imaginary part, {[}Im($\varepsilon$){]}, describes
interband optical absorption arising from electronic transitions between
occupied and unoccupied states and is directly related to the absorption
spectrum and optical bandgap. In contrast, the real part, {[}Re($\varepsilon$){]},
represents the dispersive response and determines key optical properties
such as dielectric screening, refractive index, and polarization behavior.
Notably, the static limit $\varepsilon_{\infty}$ provides insight
into the screening strength, which influences excitonic effects and
charge-carrier interactions. Together, these components offer a comprehensive
description of light-matter interaction and are essential for assessing
the suitability of these materials for optoelectronic applications.

\begin{figure}[H]

\begin{centering}
\includegraphics[width=0.48\textwidth,height=1\textheight,keepaspectratio]{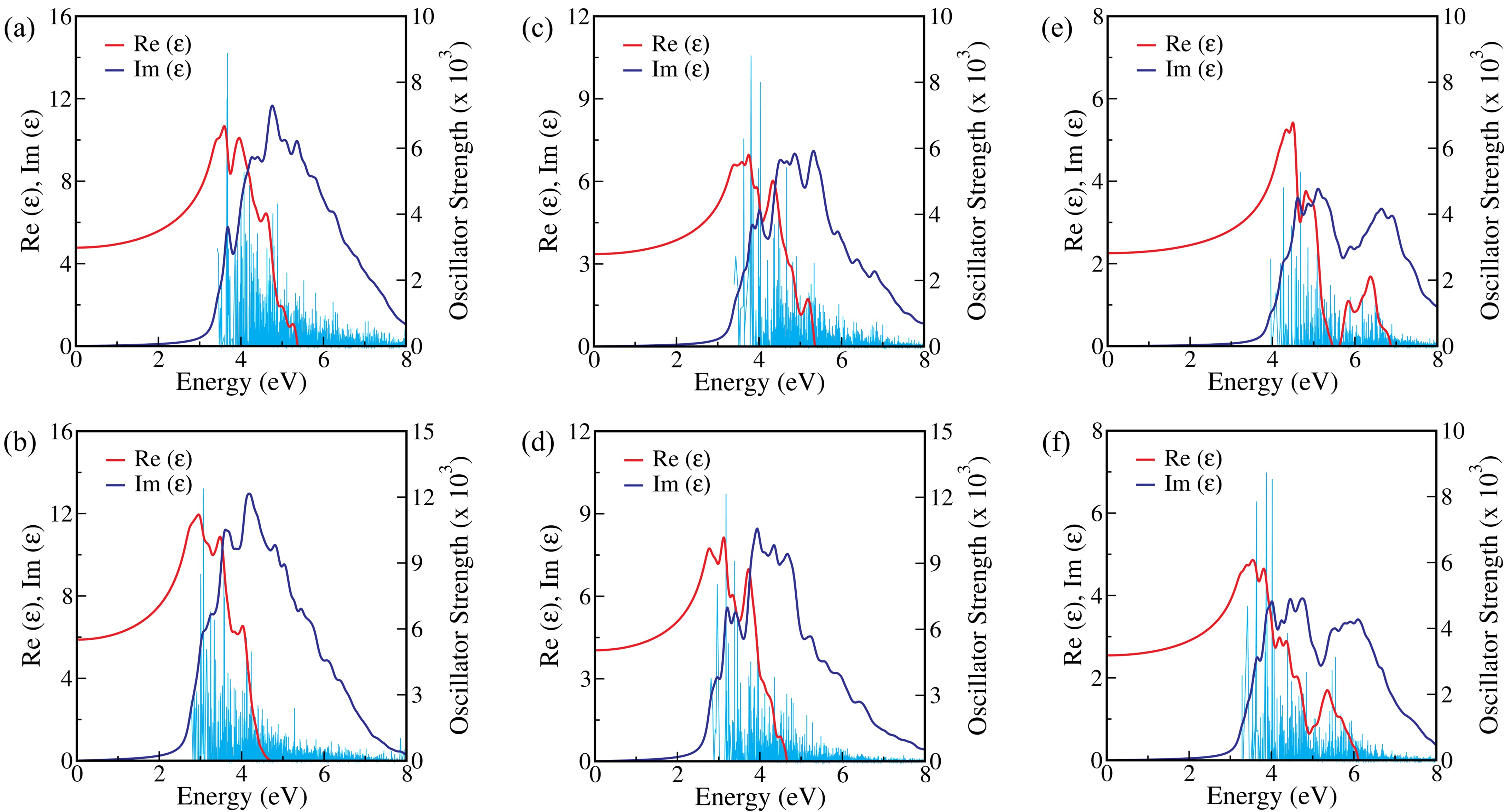}
\par\end{centering}
\caption{\label{fig:4}Spatially averaged real {[}Re($\varepsilon$){]} and
imaginary part {[}Im($\varepsilon$){]} of the electronic dielectric
function for rare-earth chalcogenide perovskites (a) YScS$_{3}$,
(b) YScSe$_{3}$, (c) LaScS$_{3}$, (d) LaScSe$_{3}$, (e) LaYS$_{3}$,
and (f) LaYSe$_{3}$, respectively, calculated using the BSE@G$_{0}$W$_{0}$@PBE
method. Peaks with cyan color represent the oscillator strength.}

\end{figure}

The imaginary part of the dielectric function, {[}Im($\varepsilon$){]},
for ABX$_{3}$ (A = Y, La; B = Sc, Y; and X = S, Se) perovskite compounds
exhibits broad spectral coverage from the visible to the ultraviolet
region, as shown in Figure \ref{fig:4}. This response reflects the
interplay of band structure, momentum selection rules, and electron-hole
interactions, where direct gaps enable sharp absorption edges, indirect
gaps lead to phonon-assisted onset, and excitonic effects renormalize
the optical gap and enhance near-edge spectral features. The
oscillator strength spectra further identify the optically allowed
excitonic transitions. Strong oscillator strength indicates a large
transition dipole moment and efficient light-matter coupling, giving
rise to intense absorption features, whereas weak or nearly vanishing
oscillator strength corresponds to optically inactive (dark) excitonic
states that contribute negligibly to the optical response. The energy eigenvalue of the first optically active (bright) exciton, which defines the optical band gap ($E_{o}$), spans 2.77$-$3.96 eV across the series, highlighting the substantial influence of excitonic effects. Since the lowest excitonic state is optically active for all compounds investigated in this work, its eigenvalue directly corresponds to the optical absorption onset. Notably, for LaScS$_{3}$,
the experimentally measured optical gap is 2.62 eV \citep{Ref-1},
whereas the present BSE optical gap value is 3.40 eV. This discrepancy
can be attributed to the idealized nature of the calculations, which
neglect temperature effects, carrier-phonon renormalization, and defect-induced
band tailing that typically reduce the measured absorption onset.
Overall, the strong and compositionally tunable optical absorption
highlights the potential of these materials for optoelectronic applications,
including photodetectors, light-emitting diodes, and laser devices.

In addition to the optical spectra, the electronic dielectric constant
($\varepsilon_{\infty}$), defined as the zero-frequency limit of
the real part of the dielectric function, is evaluated as a key descriptor
of the optoelectronic response. This quantity governs the screening
of Coulomb interactions, where larger $\varepsilon_{\infty}$ values
reduce electron-hole attraction and suppress carrier recombination,
thereby, enhancing device performance \citep{chapter2-48}. The calculated
$\varepsilon_{\infty}$ values for ABX$_{3}$ (A = Y, La; B = Sc,
Y; and X = S, Se), obtained within the BSE framework, span the range
of 2.26$-$5.88, as summarized in Table S13 of the SM \citep{supp}.
These dielectric constants also serve as essential input parameters
for evaluating excitonic and polaronic properties, as discussed in
the following sections.

\subsection{Excitonic Properties:}

In addition to the aforementioned electronic and optical properties,
excitonic properties such as, exciton binding energy ($E_{B}$), excitonic
temperature ($T_{exc}$), exciton radius ($r_{exc}$), and the probability
of the wavefunction for the electron-hole ($e-h$) pair at zero separation
($|\phi_{n}(0)|^{2}$), play a crucial role in determining the performance
of optoelectronic devices. An exciton is a Coulomb-bound $e-h$ pair
formed upon optical excitation, which renormalizes the optical gap
and governs near-edge absorption features. The exciton binding energy
($E_{B}$) quantifies the energy required to dissociate the exciton
into free charge carriers, i.e., an electron in the conduction band
and a hole in the valence band. A lower $E_{B}$ facilitates efficient
charge separation at or near room temperature, thereby promoting enhanced
photoelectric conversion efficiency, particularly in photovoltaic
applications. In contrast, a higher $E_{B}$ implies stronger Coulomb
interaction between electrons and holes, leading to more stable and
tightly bound excitons. While this can hinder efficient charge separation
and reduce photocurrent generation in photovoltaic devices, it is
advantageous for applications that rely on strong radiative recombination,
such as light-emitting diodes and laser devices, where enhanced excitonic
stability can improve emission efficiency.

The exciton binding energy ($E_{B}$) is obtained from first-principles
Bethe-Salpeter equation (BSE) calculations as, $E_{B}=E_{g}^{dir}-E_{o}$,
where $E_{g}^{dir}$ is the direct quasiparticle band gap from G$_{0}$W$_{0}$@PBE
and $E_{o}$ denotes the energy eigenvalue of the first optically active (bright) excitonic state computed within BSE@G$_{0}$W$_{0}$@PBE \citep{chapter2-38,chapter5-18}.
As summarized in Table \ref{tab:3}, the calculated $E_{B}$ values
for these chalcogenide perovskites decrease from S- to Se-containing
counterparts and lie in the range of 0.148$-$0.517 eV. Such moderately
large binding energies indicate pronounced excitonic effects arising
from reduced dielectric screening and enhanced electron-hole interactions,
which are favorable for efficient radiative recombination in light-emitting
and photodetection applications. Notably, LaYX$_{3}$ exhibits a clear
deviation from the overall trend, showing comparatively enhanced excitonic
binding. This behavior can be attributed to its reduced electronic
dielectric screening (see Figure \ref{fig:4}), which strengthens
the Coulomb interaction between electrons and holes. While the absolute
values of $E_{B}$ may be somewhat overestimated due to the use of
G$_{0}$W$_{0}$@PBE and the neglect of temperature- and phonon-induced
screening effects, the overall trends and qualitative description
of excitonic behavior remain robust. In particular, the systematic
variation across the series and the identification of intermediate
excitonic character are expected to be reliable, as they are primarily
governed by dielectric screening and carrier effective masses captured
within the present framework.

It is worth noting that when $E_{B}$ significantly exceeds the longitudinal
optical (LO) phonon energy ($\hbar\omega_{LO}$), dielectric screening
is dominated by the electronic contribution, while the ionic component
becomes negligible. Consequently, $E_{B}$ remains largely unaffected
by lattice polarization effects \citep{chapter1-65}. As shown in
Tables \ref{tab:3} and \ref{tab:5}, the condition $E_{B}\gg\hbar\omega_{LO}$
is satisfied for the present compounds, justifying the neglect of
ionic screening. This behavior is further corroborated by estimates
based on the hydrogenic Wannier-Mott model (for details, see Sec.
XII of the SM \citep{supp}). From Table S13, it is observed that
the upper bounds ($E_{Bu}$ = 0.102$-$1.009 eV) for these compounds
closely match the $E_{B}$ values obtained using the BSE@G$_{0}$W$_{0}$@PBE
method, whereas the lower bounds ($E_{Bl}$) are significantly smaller.
This indicates that the ionic contribution to the dielectric constant
is negligible, leading to $\mathrm{\varepsilon_{eff}}\rightarrow\varepsilon_{\infty}$.
Therefore, at high frequencies, dielectric screening is governed primarily
by electronic effects.

\begin{table}[H]
\caption{\label{tab:3}Calculated exciton parameters of ABX$_{3}$ (A = Y,
La; B = Sc, Y; and X = S, Se) rare-earth chalcogenide perovskites.}

\centering{}%
\begin{tabular}{cccccccccccc}
\hline 
\multirow{2}{*}{Configurations} & $E_{g}^{dir}$ &  & $E_{o}$ &  & $E_{B}$ &  & $T_{exc}$ &  & $r_{exc}$ &  & $|\phi_{n}(0)|^{2}$\tabularnewline
 & (eV) &  & (eV) &  & (eV) &  & (K) &  & (nm) &  & (10$^{27}$ m$^{-3}$)\tabularnewline
\hline 
YScS$_{3}$ & 3.747 &  & 3.424 &  & 0.323 &  & 3745 &  & 0.69 &  & 0.97\tabularnewline
YScSe$_{3}$ & 3.005 &  & 2.772 &  & 0.233 &  & 2701 &  & 1.20 &  & 0.19\tabularnewline
LaScS$_{3}$ & 3.721 &  & 3.397 &  & 0.324 &  & 3757 &  & 0.58 &  & 1.61\tabularnewline
LaScSe$_{3}$ & 2.941 &  & 2.793 &  & 0.148 &  & 1716 &  & 0.86 &  & 0.50\tabularnewline
LaYS$_{3}$ & 4.472 &  & 3.955 &  & 0.517 &  & 5994 &  & 0.32 &  & 10.14\tabularnewline
LaYSe$_{3}$ & 3.701 &  & 3.286 &  & 0.415 &  & 4812 &  & 0.43 &  & 3.87\tabularnewline
\hline 
\end{tabular}
\end{table}

Further, the calculation of $E_{B}$ within the standard first-principles
BSE framework includes only electronic screening in the $e-h$ interaction
kernel and neglects lattice (phonon) contributions. Such a static
treatment can be inadequate for polar materials, where carrier-phonon
coupling plays a crucial role in screening Coulomb interactions. To
address this limitation, we adopt the approach of Filip \textit{et
al.} \citep{chapter3-39,Ref-11}, who introduced a correction of phonon-screening
under the assumption of isotropic and parabolic band dispersion. The
correction to the exciton binding energy is given by

\begin{equation}
\Delta E_{B}^{ph}=-2\omega_{LO}\left(1-\frac{\varepsilon_{\infty}}{\varepsilon_{s}}\right)\frac{\sqrt{1+\omega_{LO}/E_{B}}+3}{\left(1+\sqrt{1+\omega_{LO}/E_{B}}\right)^{3}},
\end{equation}

where $\omega_{LO}$ denotes the characteristic longitudinal optical
phonon frequency, and $\varepsilon_{\infty}$ and $\varepsilon_{s}$
represent the electronic and static dielectric constants, respectively.
The effective $\omega_{LO}$ is evaluated using the thermal \textquotedblleft B\textquotedblright{}
approach developed by Hellwarth \textit{et al.} \citep{chapter2-22},
which accounts for the spectral contributions of multiple phonon branches
through an appropriate averaging scheme (for details, see Sec. XIII
of the SM \citep{supp}). Table \ref{tab:4} shows that phonon screening
reduces $E_{B}$ by 3.20$-$9.26\%, indicating that the overall reduction
is not substantial in these rare-earth chalcogenide perovskites. After
incorporating the phonon-screening correction, the renormalized exciton
binding energy ($E_{B}^{\prime}$) spans 0.134$-$0.498 eV. This behavior
suggests that the electronic contribution to dielectric screening
dominates over the ionic (phonon-mediated) component in these systems.

\begin{table}[H]
\caption{\label{tab:4}Calculated exciton binding energy ($E_{B}$), phonon
screening corrections ($\Delta E_{B}^{ph}$), percentage of phonon
screening contribution to the reduction of exciton binding energy
(\%), and corrected values of exciton binding energy ($E_{B}^{\prime}$)
for ABX$_{3}$ (A = Y, La; B = Sc, Y; and X = S, Se) rare-earth chalcogenide
perovskites.}

\centering{}%
\begin{tabular}{cccccccc}
\hline 
\multirow{2}{*}{Configurations} & $E_{B}$ &  & $\Delta E_{B}^{ph}$ &  & Reduction of $E_{B}$ &  & $E_{B}^{\prime}$\tabularnewline
 & (eV) &  & (meV) &  & (\%) &  & (eV)\tabularnewline
\hline 
YScS$_{3}$ & 0.323 &  & -19.20 &  & 5.94 &  & 0.304\tabularnewline
YScSe$_{3}$ & 0.233 &  & -13.55 &  & 5.82 &  & 0.219\tabularnewline
LaScS$_{3}$ & 0.324 &  & -19.41 &  & 5.99 &  & 0.305\tabularnewline
LaScSe$_{3}$ & 0.148 &  & -13.70 &  & 9.26 &  & 0.134\tabularnewline
LaYS$_{3}$ & 0.517 &  & -18.67 &  & 3.61 &  & 0.498\tabularnewline
LaYSe$_{3}$ & 0.415 &  & -13.29 &  & 3.20 &  & 0.402\tabularnewline
\hline 
\end{tabular}
\end{table}

Next, several additional excitonic parameters, including the excitonic
temperature ($T_{exc}$), exciton radius ($r_{exc}$), and the probability
of the wavefunction for the $e-h$ pair at zero separation ($|\phi_{n}(0)|^{2}$),
are evaluated and summarized in Table \ref{tab:3}. The excitonic
temperature ($T_{exc}$) serves as a useful metric to quantify the
thermal stability of bound $e-h$ pairs in semiconductors and insulators.
It is defined by relating the exciton binding energy ($E_{B}$) to
thermal energy via the Boltzmann constant ($k_{B}$) as $T_{exc}=E_{B}/k_{B}$.
Physically, $T_{exc}$ represents the characteristic temperature above
which excitons are thermally ionized into free carriers. When $T_{exc}$
significantly exceeds room temperature ($\sim$ 300 K), excitons are
expected to remain stable under ambient conditions, giving rise to
pronounced excitonic features in optical absorption and emission spectra.
Conversely, if $T_{exc}$ is comparable to or lower than room temperature,
thermal dissociation becomes efficient, and the optoelectronic response
is dominated by free carriers. In our study, the rare-earth chalcogenide
perovskite compounds exhibit high excitonic temperatures (1716$-$5994
K), reflecting strongly bound excitons that remain thermally stable
far above room temperature and yield pronounced excitonic effects
in their optical response. Such robust excitonic stability is advantageous
for optoelectronic applications, particularly in light-emitting devices
and photodetectors, where strong light-matter interaction and efficient
exciton generation are desirable.

The exciton radius ($r_{exc}$) is evaluated within the hydrogenic
Wannier-Mott model as \citep{Ref-8,Ref-9}:

\begin{equation}
r_{exc}=\frac{m_{0}}{\mu_{dir}^{*}}\mathrm{\varepsilon_{eff}}n^{2}r_{Ry},
\end{equation}

where $\mu_{dir}^{*}$ is the reduced effective mass at the direct
band edge, $\varepsilon_{\mathrm{eff}}$ denotes the effective dielectric
constant, $n$ is the exciton energy level, and $r_{Ry}$ is the Bohr
radius (0.0529 nm). In this work, the high-frequency dielectric constant
($\varepsilon_{\infty}$) is employed as the effective dielectric
screening ($\varepsilon_{\mathrm{eff}}$), and $n=1$ is considered,
corresponding to the ground-state exciton and yielding the minimum
exciton radius. The exciton radius serves as a key descriptor of the
spatial extent of the $e-h$ pair, distinguishing between localized
and delocalized excitonic states. In our study, the obtained $r_{exc}$
values ($\sim$ 3.2$-$12 $\textrm{\AA}$) indicate moderately to strongly
bound excitons with spatial extents ranging from near unit-cell localization
to a few lattice constants, suggesting an intermediate character approaching
the Wannier-Frenkel crossover regime. Such excitons promote strong
light-matter interaction, highlighting the potential of these materials
for optoelectronic applications. Within this framework, LaYX$_{3}$
stands out by exhibiting a comparatively smaller exciton radius, reflecting
enhanced localization of the electron-hole pair. This behavior originates
from its relatively weak electronic screening, which strengthens Coulomb
interactions and stabilizes more tightly bound excitonic states.

To qualitatively assess the radiative recombination
propensity of excitons, we evaluate the probability density of the
$e-h$ pair at zero separation, $|\phi_{n}(0)|^{2}$, which is given
by \citep{Ref-8,Ref-9}:

\begin{equation}
|\phi_{n}(0)|^{2}=\frac{1}{\pi(r_{exc})^{3}n^{3}}.
\end{equation}

Within the framework of radiative recombination, the
radiative decay rate is proportional to the $e-h$ overlap, i.e.,
$\tau_{rad}^{-1}\propto|\phi_{n}(0)|^{2}$. Consequently, larger values
of $|\phi_{n}(0)|^{2}$ indicate stronger $e-h$ overlap and a greater
propensity for radiative transitions. In the present case, LaYX$_{3}$
exhibits an enhanced $|\phi_{n}(0)|^{2}$, consistent with its reduced
exciton radius, confirming stronger wavefunction overlap and potentially
larger oscillator strength. Since neither radiative nor non-radiative
exciton lifetimes are explicitly calculated in this work, $|\phi_{n}(0)|^{2}$
is employed only as a qualitative descriptor of the radiative recombination
tendency rather than as a direct measure of the exciton lifetime ($\tau_{exc}$).
These results indicate pronounced light-matter interaction in the
studied systems, which is beneficial for optoelectronic and light-emitting
applications.

\subsection{Polaronic Properties:}

Understanding the fundamental limits of carrier mobility in these
chalcogenide perovskites requires a rigorous treatment of carrier-phonon
interactions within a first-principles framework \citep{chapter2-20,chapter2-21}.
In polar semiconductors, transport at ambient conditions is primarily
limited by scattering with longitudinal optical phonons, arising from
the macroscopic electric fields associated with lattice polarization
\citep{chapter5-16,chapter5-18}. This interaction leads to the formation
of polarons, wherein charge carriers become dressed by lattice distortions,
thereby altering their effective mass and transport dynamics. As a
consequence, mobility cannot be accurately described within a simple
band-like picture of free carriers.

The interaction between charge carriers and polar optical phonons
can be effectively described within the framework of the Fröhlich
Hamiltonian, which is valid in the low carrier-density regime \citep{chapter2-51}.
The strength of this interaction is quantified by the dimensionless
Fröhlich coupling constant, $\alpha$, which incorporates the effects
of dielectric screening, carrier effective mass, and the longitudinal
optical phonon frequency. It is defined as \citep{chapter5-16,chapter8-10}:

\begin{equation}
\alpha=\frac{1}{4\pi\varepsilon_{0}}\frac{1}{2}\left(\frac{1}{\varepsilon_{\infty}}-\frac{1}{\varepsilon_{s}}\right)\frac{e^{2}}{\hbar\omega_{LO}}\left(\frac{2m^{*}\omega_{LO}}{\hbar}\right)^{1/2}
\end{equation}

The calculated carrier-phonon coupling constants ($\alpha$) for the
investigated compounds are summarized in Table \ref{tab:5}. In general,
$\alpha>10$ signifies strong carrier-phonon coupling, whereas $\alpha\ll1$
corresponds to the weak-coupling regime \citep{chapter2-20}. Our
results place these materials in the intermediate-to-strong coupling
regime, with $\alpha$ values ranging from 1.92 to 10.72. Furthermore,
the carrier-phonon interaction is found to be stronger for electrons
than for holes, with this trend being particularly pronounced in LaYX$_{3}$
(X = S, Se), suggesting more significant polaronic renormalization
of electron transport. It should be emphasized that
the present analysis is based on the Fröhlich formalism, which provides
a physically meaningful description of the long-range interaction
between charge carriers and longitudinal optical phonons in polar
materials using first-principles quantities, including the dielectric
constants, effective carrier masses, and longitudinal optical phonon
frequencies. Consequently, the calculated Fröhlich coupling constants
provide useful estimates of the strength of the long-range polar electron-phonon
interaction and the associated large-polaron characteristics. However,
this approach does not explicitly evaluate the momentum- and mode-resolved
electron-phonon coupling matrix elements or identify the individual
phonon modes and electronic states contributing to the coupling. A
comprehensive microscopic treatment based on first-principles electron-phonon
coupling calculations using density-functional perturbation theory
together with Wannier interpolation would provide such information
but is beyond the scope of the present work \citep{Ref-10}. Nevertheless,
the Fröhlich model remains a widely adopted framework for describing
large-polaron behavior in polar semiconductors and is appropriate
for the objectives of the present study.

Polaron formation refers to the interaction of a charge carrier, either
an electron or a hole, with the surrounding lattice, leading to a
local structural distortion. This coupling lowers the quasiparticle
(QP) energies, so that both electron and hole states are stabilized
during polaron formation. The polaron energy, $E_{p}$, can be estimated
as follows \citep{chapter5-16,Ref-8}:

\begin{equation}
E_{p}=(-\alpha-0.0123\alpha^{2})\hbar\omega_{LO}.
\end{equation}

The QP gap associated with the polaronic states, obtained from the
electron and hole polaron energies (Table \ref{tab:5}), is compared
with the exciton binding energies, $E_{B}$, listed in Table \ref{tab:3}.
This comparison reveals that, for most of the examined chalcogenide
perovskites, the energy of the charge-separated polaronic states is
lower than that of the bound exciton states, indicating that the bound
excitons are energetically more stable. However, in the case of La(Sc,
Y)Se$_{3}$, the charge-separated polaronic states become more favorable
than the bound exciton states.

\begin{table*}[t]
\caption{\label{tab:5}Calculated polaron parameters of ABX$_{3}$ (A = Y,
La; B = Sc, Y; and X = S, Se) rare-earth chalcogenide perovskites.}

\centering{}%
\begin{tabular}{cccccccccccccc}
\hline 
\multirow{2}{*}{Configurations} & \multirow{2}{*}{$\omega_{LO}$ (THz)} & \multirow{2}{*}{} & \multicolumn{2}{c}{$\alpha$} &  & \multicolumn{2}{c}{$E_{p}$ (meV)} &  & \multicolumn{2}{c}{$m_{p}/m^{*}$} &  & \multicolumn{2}{c}{$\mu_{p}$ (cm$^{2}$V$^{-1}$s$^{-1}$)}\tabularnewline
\cline{4-5} \cline{5-5} \cline{7-8} \cline{8-8} \cline{10-11} \cline{11-11} \cline{13-14} \cline{14-14} 
 &  &  & $e$ & $h$ &  & $e$ & $h$ & \multirow{1}{*}{} & $e$ & $h$ &  & $e$ & $h$\tabularnewline
\hline 
YScS$_{3}$ & 6.50 &  & 4.39 & 2.48 &  & 124.55 & 68.79 &  & 2.21 & 1.57 &  & 2.18 & 19.04\tabularnewline
YScSe$_{3}$ & 4.86 &  & 3.99 & 1.92 &  & 84.25 & 39.56 &  & 2.06 & 1.41 &  & 2.94 & 39.94\tabularnewline
LaScS$_{3}$ & 5.92 &  & 5.56 & 3.93 &  & 145.63 & 101.01 &  & 2.70 & 2.04 &  & 2.27 & 9.67\tabularnewline
LaScSe$_{3}$ & 4.47 &  & 4.94 & 3.17 &  & 97.00 & 60.97 &  & 2.43 & 1.78 &  & 3.87 & 21.31\tabularnewline
LaYS$_{3}$ & 5.32 &  & 10.72 & 7.04 &  & 267.32 & 168.53 &  & 5.66 & 3.41 &  & 0.21 & 2.12\tabularnewline
LaYSe$_{3}$ & 3.87 &  & 10.58 & 6.30 &  & 191.63 & 108.79 &  & 5.56 & 3.04 &  & 0.34 & 5.05\tabularnewline
\hline 
\end{tabular}
\end{table*}

Feynman introduced a powerful variational approach to address the
Fröhlich Hamiltonian, which describes the interaction between an electron
(or hole) and a continuum of independent, harmonically oscillating
phonon modes within a quantum field theoretical framework \citep{chapter2-23}.
As the electron (or hole) moves through the lattice, it interacts
with the polarization field it induces, which evolves and decays over
time. Within this formalism, and in the weak-coupling limit (small
$\alpha$), the polaron effective mass, $m_{p}$, can be expressed
as \citep{chapter2-23,Ref-PRB}:

\begin{equation}
m_{p}=m^{\ast}\Big(1+\frac{\alpha}{6}+\frac{\alpha^{2}}{40}+...\Big).
\end{equation}

As shown in Table \ref{tab:5}, carrier-phonon coupling leads to a
substantial enhancement of the effective mass, with $m_{p}$ increasing
by approximately 41$-$466\%. This significant renormalization indicates
the presence of intermediate to strong carrier-lattice interactions
in the examined systems.

To further assess the impact of the enhanced polaron effective mass,
the polaron mobility is estimated using the Hellwarth polaron model
as \citep{chapter2-22,Ref-9}:

\begin{equation}
\mu_{p}=\frac{\left(3\sqrt{\pi}e\right)}{2\pi c\omega_{LO}m^{*}\alpha}\frac{\sinh(\beta/2)}{\beta^{5/2}}\frac{w^{3}}{v^{3}}\frac{1}{K(a,b)}
\end{equation}

where $e$ is the electronic charge, $\beta=hc\omega_{LO}/k_{B}T$,
$w$ and $v$ are temperature-dependent variational parameters, and
$K(a,b)$ is a function of $\beta$, $w$, and $v$ (for details,
see Sec. XIII of the SM \citep{supp}). The polaron mobility quantifies
the ease with which a polaron propagates through the lattice and is
governed by both its effective mass and the strength of carrier-lattice
interactions. As the effective mass increases due to strong carrier-phonon
coupling, the mobility correspondingly decreases, leading to slower
carrier transport. This behavior is evident in the present systems
(Table \ref{tab:5}), where hole polarons exhibit significantly higher
mobilities (2.12$-$39.94 cm$^{2}$V$^{-1}$s$^{-1}$) compared to
electron polarons (0.21$-$3.87 cm$^{2}$V$^{-1}$s$^{-1}$). This
indicates stronger electron-phonon coupling and/or larger effective
masses for electrons, leading to more localized electron polarons
and consequently reduced mobility. In contrast, the relatively lighter
and less strongly coupled hole polarons enable more efficient charge
transport. Despite the relatively low electron mobility, the favorable
hole transport, combined with the structural stability and non-toxicity
of these materials, underscores their potential for optoelectronic
applications, particularly in devices where hole conduction plays
a dominant role.

A broader perspective can be obtained by comparing the present III-III
chalcogenide perovskites with the more extensively studied II-IV systems
ABX$_{3}$ (A = Ca, Sr, Ba; B = Ti, Zr, Sn, Hf; and X = S, Se). The
latter typically exhibit smaller band gaps ($\sim$ 1.0$-$2.5 eV),
larger dielectric screening, and correspondingly weaker excitonic
effects, resulting in more delocalized Wannier-type excitons and longer
carrier diffusion lengths that are advantageous for photovoltaic applications
\citep{chapter1-63,chapter3-19,chapter5-16,chapter6-9,Ref-PRB}. In
addition, relatively weaker carrier-phonon coupling in these systems
leads to lighter polarons and higher carrier mobility. In contrast,
the III-III compounds investigated here display wider band gaps (2.75$-$4.47
eV), reduced dielectric screening, and moderately large exciton binding
energies, giving rise to more localized excitons with enhanced electron-hole
overlap. This enhanced localization is further accompanied by stronger
carrier-phonon coupling, leading to heavier polaronic quasiparticles
and reduced carrier mobility. While such polaronic effects may limit
efficient charge transport and charge separation, they can simultaneously
stabilize excitonic states and suppress non-radiative recombination.
Consequently, the combination of strong excitonic binding and polaronic
localization results in pronounced light-matter interaction and enhanced
radiative recombination, making these materials particularly promising
for light-emitting and photodetection applications, albeit less favorable
for photovoltaic performance.

\section{Conclusion:}

In summary, we have systematically investigated the ground- and excited-state
properties of III-III rare-earth chalcogenide perovskites ABX$_{3}$
(A = Y, La; B = Sc, Y; and X = S, Se) using state-of-the-art density
functional theory, density functional perturbation theory, and many-body
perturbation theory. The calculated phonon band structures and elastic
constants confirm the dynamical and mechanical stability of these
compounds. Our electronic structure analysis reveals quasiparticle
band gaps (G$_{0}$W$_{0}$@PBE) in the range of 2.75$-$4.47 eV,
accompanied by lower hole effective masses compared to electrons,
indicating favorable p-type transport. The optical properties, computed
within the BSE framework, exhibit a strong absorption onset spanning
the visible to ultraviolet region, highlighting their potential for
optoelectronic applications. Importantly, these compounds exhibit
pronounced excitonic effects, hosting stable excitons as evidenced
by their intermediate-to-large binding energies (0.148$-$0.517 eV),
high excitonic temperatures, Wannier-Frenkel crossover character,
and strong electron-hole wavefunction overlap, indicative
of pronounced light-matter interaction and favorable radiative recombination
characteristics. Fröhlich\textquoteright s mesoscopic analysis indicates
appreciable carrier-phonon coupling in these systems, with electron-phonon
interactions consistently outweighing their hole counterparts. Remarkably,
bound excitonic states dominate over charge-separated polaronic configurations
in nearly all compounds, with La(Sc, Y)Se$_{3}$ as an exception,
while hole polarons demonstrate substantially higher mobilities (reaching
up to $\sim$ 40 cm$^{2}$V$^{-1}$s$^{-1}$) than electrons. Overall,
the interplay between strong excitonic effects and favorable polaronic
transport, combined with excellent structural stability and lead-free
composition, establishes rare-earth chalcogenide perovskites ABX$_{3}$
as promising candidates for next-generation optoelectronic technologies.
These materials are particularly attractive for applications in light-emitting
devices and photodetectors, and they offer a viable platform for exploring
excitonic optoelectronics in environmentally benign systems.
\begin{acknowledgments}
The authors would like to acknowledge the Council of Scientific and
Industrial Research (CSIR), Government of India {[}Grant No. 3WS(007)/2023-24/EMR-II/ASPIRE{]}
for financial support. The authors acknowledge the High Performance
Computing Cluster (HPCC) \textquoteleft Magus\textquoteright{} at
Shiv Nadar Institution of Eminence for providing computational resources
that have contributed to the research results reported within this
paper.
\end{acknowledgments}

\section*{DATA AVAILABILITY}

The data that support the findings of this article are not publicly
available. The data are available from the authors upon reasonable
request.

\bibliography{refs}

\end{document}